\documentstyle[11pt,IAU207_pasp,twoside,psfig]{article}
\begin{document}

\title{The Outer Halo Cluster System of NGC 1399}

\author{B. Dirsch, D. Geisler, T. Richtler}
\affil{Universidad de Concepci\'on, Chile}
\author{J.C. Forte}
\affil{Universidad Nacional de La Plata, Argentina}

\begin{abstract}

We investigate the outer halo globular cluster population of NGC 1399. 
This study uses wide-field imaging of this cluster system, which covers 
the largest area studied with CCD photometry until now. 
The cluster system of NGC 1399 is found to extend further than
100 kpc from the galaxy. A population of metal-rich, as well as metal-poor
clusters has been
identified at these large radii.  At radii smaller than 55 kpc the specific 
frequency of the red cluster system remains constant, while that of the 
blue clusters increases proportional to $r^{0.8\pm0.2}$. For larger radii,
the uncertainty of the galaxy light profile does not permit any reliable
statement.

\end{abstract}

\section{Introduction}

NGC 1399, the central galaxy of the Fornax cluster, is, besides M87, the 
nearest central cD galaxy in a relatively dense cluster. 
Hanes \& Harris (1986) already noted in an earlier photographic work that the cluster system
extends further than $13\arcmin$, which corresponds to a distance of
70 kpc at the distance of NGC 1399 (assumed to be 19 Mpc). While the properties
of the inner cluster system are well known (Ostrov et al. 1998), the population
structure of the outer cluster system was virtually unknown. New wide-field CCD
cameras make studies of this extended system now feasible.

\section{Observations \& Reduction}

MOSAIC wide-field images ($36\arcmin\times 36\arcmin$) in the Washington $C$ 
and the Kron-Cousins $R$ filter
have been obtained at the CTIO 4m telescope centered near NGC 1399 (in Dec. 1999). 
In addition a background field 3.5 degrees north-east of NGC 1399 has 
been observed.  We used SExtractor 
and DAOPHOT for identification and photometry of the objects, respectively.
Later spectroscopic observations with the VLT showed that the success rate for
the identification of GCs between $2\arcmin$ and $8\arcmin$ based on this
photometry was higher than 95\% (Richtler et al. 2001). In total, we identified
about 5000 cluster candidates out to a radial distance of $22\arcmin$ and down to
a limiting magnitude of R=24.

\section{The different cluster populations}

For the current study we selected only
clusters brighter than $R=23$, well above the completeness limit.
The globular cluster candidates were divided into blue clusters ($1<C-R<1.4$) 
and  red clusters ($1.5<C-R<2.2$).  
The reason for doing that is the strong color bimodality
of the cluster system. The color difference  of $C-R = 0.47$
between the peak values of
these two populations is most probably due to a mean metallicity difference 
of approximately 1\,dex.
Bimodal distributions are typical for large
cluster galaxies as shown by Gebhardt \& Kissler-Patig (1999)
and Forbes \& Forte (2001). 
The mean metallicity values for the two 
populations are [Me/H]$= -0.3\pm 0.15$ and [Me/H]$= -1.3\pm0.15$. 

The two populations have rather different spatial distributions with the 
blue clusters being less concentrated than the redder ones, as shown 
in Fig.1. Moreover, each population changes its power-law 
index at about $7\arcmin$.
We find surface density profiles for the blue population proportional to
$r^{-0.8\pm0.15}$ and $r^{-1.4\pm0.14}$ (for the inner and outer part, respectively), 
and for the red population profiles proportional to
$r^{-1.5\pm0.15}$ and $r^{-1.8\pm0.18}$.  

\begin{figure}[t]
 \centerline{\vbox{\psfig{figure=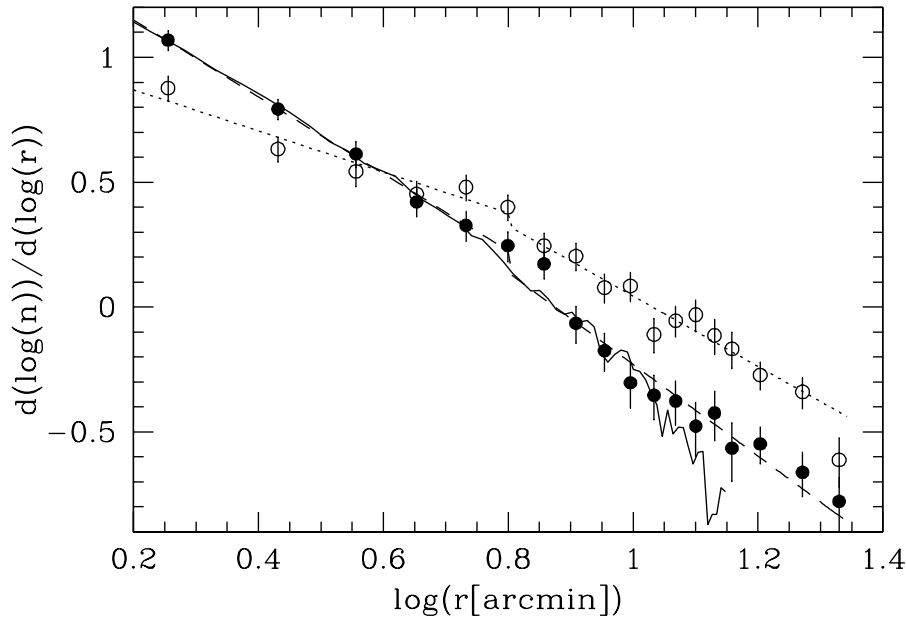,width=12cm}}}
 \caption{The radial density profiles of red (filled circles) 
	and blue (open circles) globular clusters are plotted together with the fitted power-laws.
	The solid line shows the galaxy light profile in the R filter.  }
 \label{fig:GCslope}
\end{figure}

The galaxy light profile measured by us in the $R$ band closely matches the radial 
distribution of the red cluster population within $11\arcmin$.
Outside of $11\arcmin$ small background uncertainties make a definite statement
on the surface brightness profile impossible.
The blue cluster system on the other hand
has within $11\arcmin$ a shallower profile and therefore its  specific frequency 
with respect to the R band luminosity
increases outwards proportional to $r^{0.8\pm0.2}$.

The $B$ band surface brightness profile determined by Caon et al. (1994) and
our profile are in excellent agreement, which indicates that no 
color gradient in $B-R$ is visible within $11\arcmin$. Together with the fact
that the red clusters closely follow the galaxy light, the question remains
why the dominance of the blue clusters at larger radii is apparently not accompanied 
by a color gradient in the galaxy light. However, more accurate photometry at 
low surface brightness levels is required to resolve this issue.

To investigate the very remote cluster population, 
we selected all cluster candidates which are further away than $10\arcmin$ from any
bright Fornax galaxy and more than $13\arcmin$ away from NGC 1399. 
The blue peak of the color distribution of these remote clusters has the same color
and the same width as the blue peak of the inner blue clusters.
A red 
peak on the other hand cannot unambiguously be identified; however, a shoulder at
the position of the red peak of the inner population can be seen. We estimate 
the mean metallicity of this red cluster population to be roughly $-0.6$ dex.
The similarity of the color distribution of the remote blue clusters
suggests that they are still associated with the outer halo of NGC 1399, indicating that its 
cluster system extends to beyond 100 kpc. An analogous statement for the red clusters is more difficult. 

It remains an intriguing question whether these clusters and the ''intergalactic'' 
clusters observed
by Bassino et al. (2001) are connected. They found a population of clusters 
at large distances from the center from the Fornax cluster that have a very similar 
metallicity distribution as our remote cluster sample. One might speculate that
this population is a tracer of a global stellar population spread out over the Fornax cluster.

\section*{Acknowledgments}
We acknowledge the help of M.Shara, D.Zurek and E.Grebel, who obtained some images that were
used in this investigation.

\section*{References}
\begin{quote}
Bassino L., Forte J.C., 2001, IAU Symp. 2001, this volume\\ 
Caon N., Capaccioli M., DÓnofrio M., 1994, A\&AS 106, 199\\
Forbes D.A., Forte J.C., 2001, MNRAS 322, 257 \\
Gebhardt K., Kissler-Patig M., 1999, AJ 118, 1526\\
Hanes D.A., Harris W.E., 1986, ApJ 309, 564\\
Ostrov P.G, Forte J.C., Geisler D., 1998, AJ 116, 2854 \\
Richtler T. et al. 2001, IAU Symp. 2001, this volume\\
\end{quote}

\end{document}